\documentclass[12pt]{article}
\usepackage{amssymb,amsmath,graphicx,latexsym}
\begin{document}
\title{\textbf{NLO tools and the background to $H\rightarrow WW$}} 
\author{B.~Holdom\thanks{bob.holdom@utoronto.ca}\\
\emph{\small Department of Physics, University of Toronto}\\[-1ex]
\emph{\small Toronto ON Canada M5S1A7}}
\date{}
\maketitle
\begin{abstract}
A theory input is needed for the estimation of the largest background for the $H\rightarrow WW$ search at the LHC. This is the shape of the $m_{\ell\ell}$ spectrum from continuum $WW$ production. We find this to depend on how NLO matrix elements are merged with parton showers, and we compare the results from a number of different implementations. The results suggest that both the size of the background estimate and its uncertainty may have been underestimated. This conclusion is reinforced by a ``Note Added'', which comments on the LHC results released on November 14 2012.
\end{abstract}

The Higgs to $WW$ search channel at the LHC suffers from substantial backgrounds, with the dominant one being continuum $WW$ production. This background in the signal region is estimated by counting events in a control region, subtracting off other backgrounds to deduce the number of continuum $WW$ events in the control region, and then extrapolating that number to the signal region. This extrapolation requires theory input, namely the ratio of $WW$ cross sections in the signal and control regions. This ratio is labeled $\alpha$ in \cite{diff} and it is obtained from a particular NLO Monte Carlo simulation in the case of ATLAS \cite{atlas} and from a unspecified simulation in the case of CMS \cite{cms}. Since more details are given by ATLAS we shall focus on their analysis of the decay mode $WW\rightarrow e\nu\mu\nu$. We shall also focus on the 0-jet bin and on the ratio $\alpha_0$. In this bin the $WW$ background is about 70\% of the total background in the signal region. A 125 GeV Higgs signal is about 14\% of the $WW$ background (here NNLO effects are included in the Higgs signal estimation), and thus theoretical uncertainties in $\alpha_0$ of similar size can have a relatively large impact on the analysis.

In terms of a fixed order parton level description $\alpha_0$ can be obtained at NLO. A calculation of $\alpha_0$ at NNLO may eventually become available since a complete NNLO calculation of diphoton production has been performed \cite{diphoton} and partial NNLO results for $WZ$ production have also appeared \cite{WZ}. In these examples the NNLO corrections are large, but currently the potential impact of such corrections for the $H\rightarrow WW$ background have not been considered. For the simulation of background at the detector level fully showered and hadronized events are required, and so at the very least parton showers need to be merged with the NLO event generation. As we shall explore, the introduction of parton showering tends to increase $\alpha_0$ from the NLO result, and as well the different implementations of the parton shower give a range of values. The parton shower includes some effects that would be included at NNLO, and thus these results can also give some idea of the possible size of higher order corrections.

The value of $\alpha_0$ used in the ATLAS analysis can be obtained from Table 2 of  \cite{atlas} as follows.\footnote{Pierre Savard, private communication.} The listed number $WW$ of events in the signal region after cuts (234) includes a correction factor. This factor when applied to the listed number of $WW$ events in the control region (531) is designed to push the total number of events in the control region (789) from all backgrounds up to the observed value (820). Thus $\alpha_0=234/(531+820-789)=0.416$. We emphasize that $\alpha_0$ is not determined from data, that a Monte Carlo simulation is needed to obtain it, and that it determines the overall number of $WW$ background events in the signal region given measurements in the control region.

We shall define another quantity $\alpha_0'$ which is slightly different from $\alpha_0$. It has some advantages: 1) it is a property of a single distribution, the $m_{\ell\ell}$ distribution in the presence of certain cuts, 2) it does not depend on Monte Carlo modeling of the tail of this distribution at arbitrarily high $m_{\ell\ell}$, 3) it can be obtained from \cite{atlas} without ambiguity, namely from Fig.~(14b). We define $\alpha_0'$ as the ratio of the number of continuum $WW$ events in the $10<m_{\ell\ell}<50$ GeV region over the number in the $80<m_{\ell\ell}<290$ GeV region with the cuts that were used to obtain Fig.~(14b). 290 GeV is the maximum $m_{\ell\ell}$ in Fig.~(14b). By discretizing the figure and pulling out the $WW$ contribution we obtain $\alpha_0'=0.457$. $\alpha_0'$ is 10\% larger than $\alpha_0$ and this difference is due to 1) not including the $\Delta\phi_{\ell\ell}<1.8$ cut on the signal region, and 2) omitting the contribution from $m_{\ell\ell}>290$ GeV in the control region. Of the 10\% difference, from Table 2 in \cite{atlas} we see that the first effect gives 4\%, and so the $m_{\ell\ell}>290$ GeV tail is a 6\% effect according to ATLAS.

For the $WW$ background ATLAS \cite{atlas} uses MC@NLO and Herwig6 for event generation followed by a full detector simulation. We shall obtain $\alpha_0'$ from a number of different NLO Monte Carlo tools, but without the detector simulation. The range of these values will translate into a theoretical uncertainty in the final estimate of the $WW$ background.

Our representation of the ATLAS cuts is as follows.
\begin{itemize}
\item  a $e^\pm\mu^\mp$ pair, with $p_T$ thresholds of 25 and 15 GeV for the two leptons
\item $|\eta|<2.5$ for leptons
\item $\Delta R>0.3$ between leptons
\item $p_T^{\ell\ell}>30$ GeV
\item no jet with $p_T>25$ GeV (anti-$k_T$ algorithm with $R=0.4$)
\item $E^{\rm miss}_T\sin(\min(\Delta\phi,\pi/2))>25$ GeV
\end{itemize}
$\Delta\phi$ is the minimum azimuthal angle between ${\bold E}_T^{\rm miss}$ and either of the two leptons. With these cuts our values of $\alpha_0'$ are obtained as the ratio of cross sections in the $10<m_{\ell\ell}<50$ GeV and $80<m_{\ell\ell}<290$ GeV regions. Sometimes we will replace the jet veto with another nearly equivalent cut described below. $\Delta R>0.3$ is only included because it is effectively implied by lepton isolation requirements, and if omitted $\alpha_0'$ would only increase very slightly.

We apply these cuts at the generator level while ATLAS applies these cuts at the detector level to obtain their Fig. (14b). At the detector level lepton isolation requirements are imposed and many other factors come into play, such as jet scale uncertainties, the modeling of pile-up, triggering efficiencies, etc. We shall deduce that detector level effects cause a substantial decrease in $\alpha_0'$, or in other words a significant distortion of the $m_{\ell\ell}$ distribution. The large size of the detector level effects suggest that their modeling could be another source of uncertainty.

Before considering the $m_{\ell\ell}$ distribution we first consider another leptonic distribution which is especially sensitive to the introduction of the parton shower. The $p_T$ of the $WW$ system is obviously sensitive to the partonic and gluonic degrees of freedom against which it recoils. $p_T^{WW}<25$ GeV is the relevant range for the 0-jet bin and we display examples of this distribution in the presence of the cuts in Fig.~(1). The POWHEG-BOX \cite{powheg} package allows direct comparison of the fixed order NLO result and the LHE result (from the Les Houches event file before showering) with the showered results using Pythia6 \cite{pythia} and Herwig6 \cite{herwig6}. We also include the result from MC@NLO (4.09) \cite{mcnlo} using Herwig6 for showering. The POWHEG LHE events include a (N)LL resummation of soft gluon effects. We see that this is the key feature of the parton shower which removes the infrared singularity of the fixed order NLO result and completely changes the $p_T^{WW}$ spectrum in the 0-jet bin.
\begin{figure}[t]
\centering\includegraphics[scale=.8]{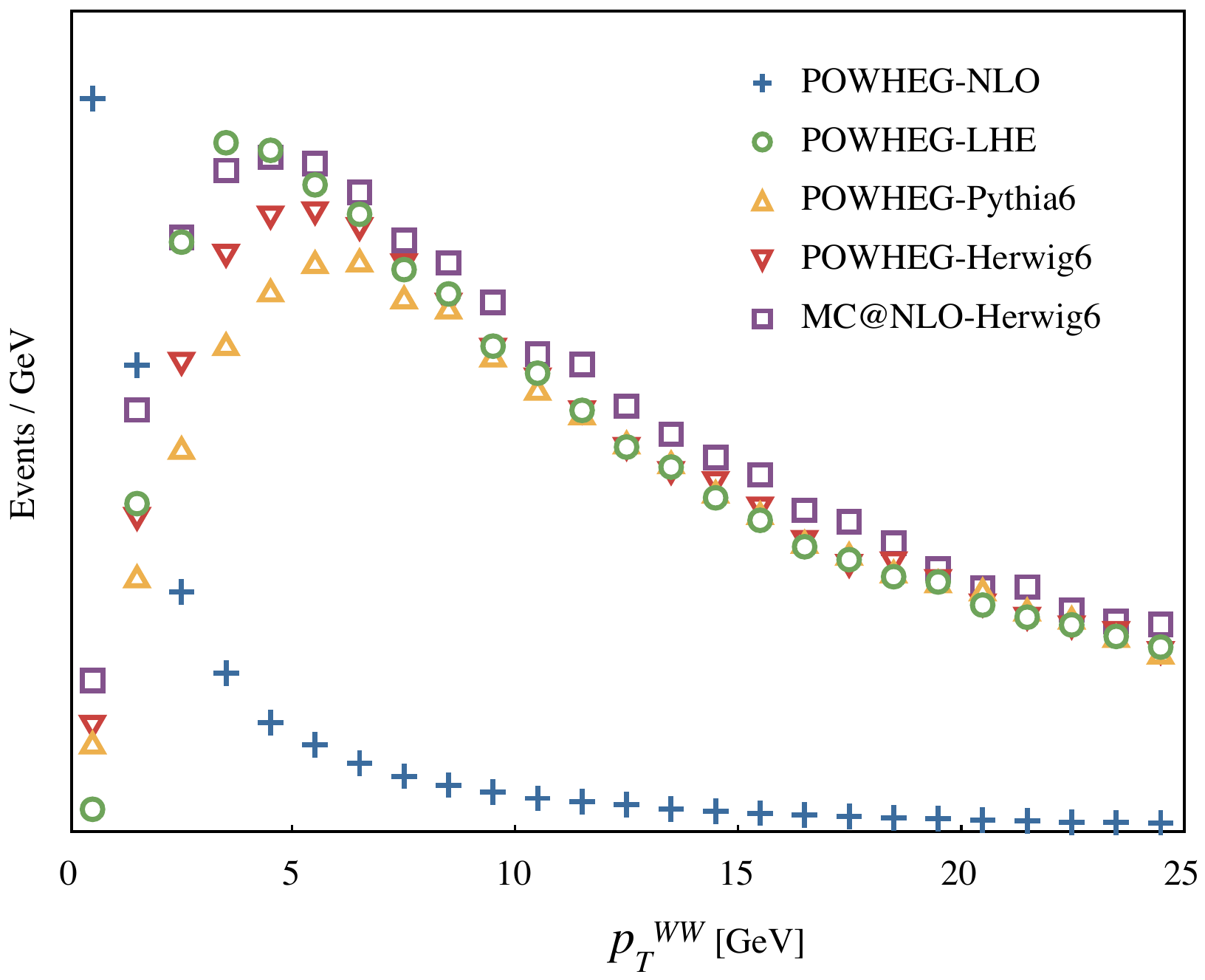}
\caption{The $p_T^{WW}$ distribution. The NLO result has a singularity structure at $p_T=0$ such that the actual value for the lowest bin is the negative of the value shown. The NLO values are also divided by 14.}\end{figure}

MC@NLO and POWHEG generated events can also be showered through the newer programs Herwig++ (2.6.1) \cite{herwig} and Pythia8 (8.17) \cite{pythia8} respectively. Another independent result can be obtained from Herwig++ which has its own internal implementation \cite{hamilton} of a POWHEG-style NLO event generator. We label this tool Herwig++@NLO. We also include a result from a beta version of aMC@NLO \cite{amc} available in the MadGraph 5 framework \cite{mad}. To implement a common analysis for these newer tools, and for Sherpa (1.4.2) \cite{sherpa} as well, we make use of their ability to generate HepMC \cite{hepmc} format event files. This is in contrast to the tools mentioned in the previous paragraph, where we instead use their built-in analysis routines suitably modified to include the cuts. We also use the strict NLO parton level tools, MCFM (6.3) \cite{mcfm} and VBFNLO (2.6.0) \cite{vbfnlo}, with their built-in analysis routines.

When available we feed the HepMC event files through the Delphes fast detector simulator \cite{delphes}.\footnote{Delphes is slightly modified to properly handle the negative weights in the HepMC events produced by MC@NLO and aMC@NLO.} This implements jet finding based on FastJet \cite{fastjet}, where we select the anti-$k_T$ algorithm with $R=0.4$. Delphes produces LHCO files containing the reconstructed objects, and it is at this stage that we apply the cuts listed above via MadAnalysis \cite{mada}. We turn off lepton isolation requirements and the trigger in Delphes to minimize detector effects. The tracking efficiency is found to have no impact on $\alpha_0'$ and we set it to 100\%. The remaining aspects of detector simulation that Delphes still implements are found to have very little impact on the value of $\alpha_0'$.
\begin{table}[t]
\centering
\begin{tabular}{c|c|c}ATLAS & 0.447 \\\hline\hline Sherpa-LO & 0.54 \\\hline Sherpa-LO-shower & 0.58\\\hline\hline MCFM & 0.555$^\#$ \\\hline VBFNLO & 0.555$^\#$ \\\hline POWHEG-NLO & 0.55 \\\hline POWHEG-LHE & 0.57 \\\hline\hline MC@NLO-Herwig6 & 0.55 \\\hline POWHEG-Pythia6 & 0.59 \\\hline POWHEG-Herwig6  & 0.60\\\hline\hline aMC@NLO-Herwig++ & 0.545 \\\hline MC@NLO-Herwig++ & 0.55 \\\hline POWHEG-Pythia8 & 0.59 \\\hline Herwig++@NLO & 0.625 \end{tabular}
\caption{Values of $\alpha_0'$. A number of runs are compared in each case to arrive at standard error estimates less than $\pm0.005$. The ATLAS number includes the effect of detector simulation while the other numbers do not. $^\#$These numbers include a 3\% contribution from $gg\rightarrow WW$.}\end{table}

We thus obtain the $m_{\ell\ell}$ distributions with the cuts imposed, and from these we obtain the values for $\alpha_0'$ listed in Table 1. We have used Sherpa to obtain the lowest order result for $\alpha_0'$. We also use Sherpa to introduce a parton shower through its CKKW style merging of the shower with matrix elements involving 0 and 1 additional partons. Sherpa in addition implements an internal POWHEG-style NLO event generator, but currently it is unable to produce unweighted events for NLO processes and so this precludes an NLO entry from Sherpa.

As we have mentioned the similar quantity $\alpha_0$ was introduced and studied in \cite{diff}. No numerical values of $\alpha_0$ were given but it was noted that MCFM and MC@NLO-Herwig6 produced values in good agreement. The suggestion from this reference is that the theoretical errors on $\alpha_0$ are small. We also see the agreement between MCFM and MC@NLO-Herwig6, but the other results in Table 1 suggest a different conclusion regarding theoretical errors.

Both MCFM and VBFNLO implement the $gg\rightarrow WW$ process which proceeds through a quark loop. Although strictly of higher order it is numerically relevant due to enhancement from the gluon PDFs. ATLAS also incorporates this process but the other event generators in Table 1 do not implement it. VBFNLO allows $\alpha_0'$ to be obtained with and without this contribution, and it is found that the $gg\rightarrow WW$ process increases $\alpha_0'$ by about 3\%. Thus to include this effect the results from the other generators should be increased by 3\%.

MC@NLO-Herwig6 is the event generator used by ATLAS \cite{atlas} (but see Note Added), and we see that the value it gives for $\alpha_0'$ is interesting for two reasons. This $\alpha_0'$ happens to be at the low end of the NLO+shower estimates, and so in this sense the ATLAS estimate of the $WW$ background could be on the low side. And second this value for $\alpha_0'$ is about $(20+3)\%$ larger than the $\alpha_0'$ extracted from the ATLAS analysis. This surprisingly large difference is the amount that detector level effects have distorted the $m_{\ell\ell}$ distribution. A distortion of this size in a simple leptonic distribution, which happens to be in the direction of lowering the apparent background, deserves further study.

Table 1 displays some interesting patterns among the showered results. The source of the differences is mainly due to the different NLO+shower implementations rather than the different generators used for showering. The MC@NLO (and aMC@NLO) values are similar to the pure NLO value, while the POWHEG values are larger. In particular when MC@NLO and POWHEG are compared with the same parton shower (Herwig6) the $\alpha_0'$ from POWHEG is $\approx9\%$ larger.  The Herwig++@NLO implementation yields an even larger $\alpha_0'$ that is $\approx14\%$ larger than the MC@NLO value. This is a significant difference since it is larger than the size of the Higgs signal.

The differences between ATLAS, MC@NLO-Herwig6 and Herwig++@NLO are made clear in Fig.~(2). The $m_{\ell\ell}$ distributions have been normalized to have an equal weight in the $80<m_{\ell\ell}<290$ GeV region. The $10<m_{\ell\ell}<50$ GeV signal region is indicated.

\begin{figure}[t]
\centering\includegraphics[scale=.8]{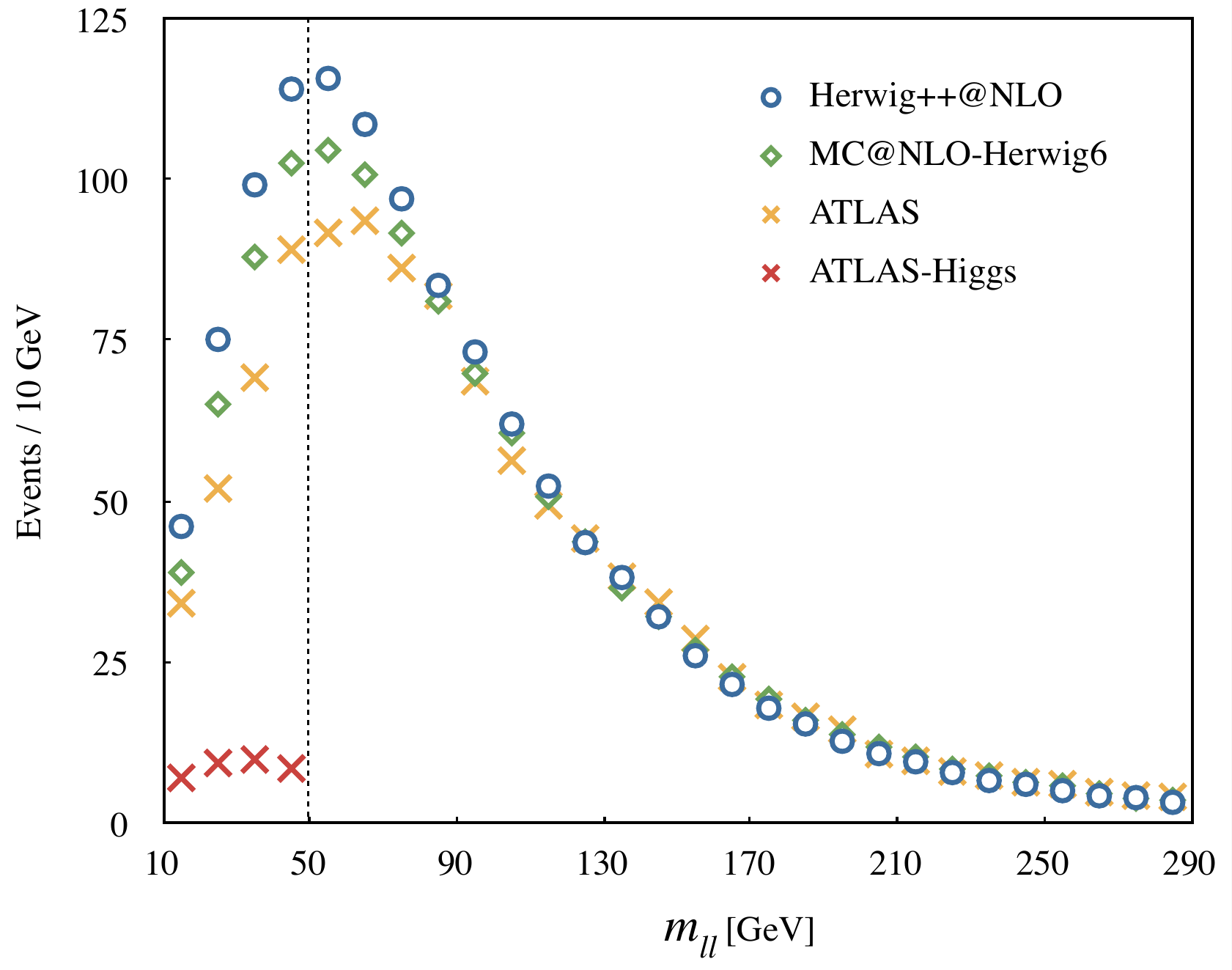}
\caption{Comparison of $m_{\ell\ell}$ distributions of the $WW$ continuum background normalized in the $80<m_{\ell\ell}<290$ GeV region. Only the ATLAS values include the full detector simulation. The two event generators do not include the $gg\rightarrow WW$ contribution. Also shown is the simulated 125 GeV Higgs signal in the signal region. ATLAS results are obtained from Fig. (14b) \cite{atlas}.}\end{figure}

We now give some details that could be useful for reproducing our results. At the strict NLO level the $WW$ system recoils against a single extra parton. Thus we could expect that replacing the jet veto with a cut $p_T^{WW}<25$ GeV would produce very similar results. We confirm that this is true at the level of showered events by using POWHEG-BOX, which implements a FastJet algorithm on its showered events. Thus to avoid problems associated with jets in the various analysis routines, for example when events are not showered or when jet finding is not available, we use the $p_T^{WW}$ cut rather than the jet veto in all cases where Delphes is not used.

MC@NLO, aMC@NLO and Herwig++@NLO are run with their default choices for the renormalization scale while the renormalization scale for POWHEG, Sherpa, MCFM and VBFNLO is chosen to be the mass of the $WW$ system, $m_{WW}$. Typically we find that the choice of renormalization scale has only a mild effect on $\alpha_0'$, at least compared to the other variations of $\alpha_0'$ we are finding.

The cteq6l1 PDF is used with Sherpa while the CT10 PDF is used in nearly all other cases, both at the NLO event generation stage and in the showering stage. The exception is the POWHEG-BOX package where CT10 is used for event generation but Pythia6 and Herwig6 are used with their default PDFs. We have not considered the uncertainties associated with PDFs. For the studies involving HepMC event files we turn off the hadronization and multiple parton interactions since we find that this has little or no impact on $\alpha_0'$ (unless we consider lepton isolation cuts, see below). Pythia6 and Herwig6 are run with hadronization turned on. 

One might wonder about leptonically decaying $\tau$'s originating from one or both of the $W$'s. We have used Sherpa to check that these configurations by themselves give a $\alpha_0'$ just slightly larger than usual. So this effect on $\alpha_0'$ is negligible.

Other than what we have described the various programs are run essentially with their default settings, and any further details can be obtained from the author. We mention in particular that Pythia8 gives the user easy access to a number of settings for the showering of POWHEG events; a sampling of different choices for these settings gives values of $\alpha_0'$ that differ by a few percent.

We return to the question of the detector level effects, which we have thus far avoided in our use of Delphes. To see what a fast detector simulator can say about these effects we first turn on lepton isolation cuts as follows. For both electrons and muons the summed $p_T$ of tracks in a $R=0.3$ cone around the lepton (excluding the lepton itself) is required to be less than 0.1. For muons the ratio of $E_T$ in a $3\times3$ calorimeter array around the muon (including the muonÕs cell) to the $p_T$ of the muon is required to be less than 0.1. When Delphes is run on fully showered and hadronized events with MPI turned on, we find a 5\% reduction in the value of $\alpha_0'$. So this goes a little ways to bridge the gap between our values of $\alpha_0'$ and the ATLAS value.

When we turn the Delphes trigger emulation on we find that this has no influence on $\alpha_0'$. This is not surprising since the same information is used in Delphes for the trigger emulation as for the final analysis, and so Delphes is blind to the differing resolutions inherent in the real triggers. The real triggers could in principle affect $\alpha_0'$.

Another quantity for which a fast detector simulator is over idealized is $E_T^{\rm miss}$, and so we briefly consider the effect of deviations between true and measured $E_T^{\rm miss}$. We can do this in the final analysis of the LHCO events, where to each event we add a fake missing energy vector of some fixed magnitude and random direction in the transverse plane. This will modify the effect of the missing energy cut described above. We find that the effect is to \textit{increase} $\alpha_0'$ by a percent or two for 20 GeV of fake missing energy per event. So the answer does not appear to lie with missing energy. 

To summarize, the shape of the $m_{\ell\ell}$ distribution from continuum $WW$ production is crucial for estimating the background for the $H\rightarrow WW$ search. We find that the effect of merging parton showers with NLO event generation can cause greater enhancement of the background estimate than caused by the strictly NLO corrections. There is  a significant difference in the amount of enhancement depending on whether MC@NLO, POWHEG or Herwig++@NLO is used. This translates into a significant theoretical uncertainty which may not have been fully accounted for in the experimental analyses. Large detector level effects also remain obscure.

There is no guarantee that the sampling of results from different tools fully reflects the true theoretical uncertainty. One could hope for guidance from the measurements of the $WW$ cross sections at the LHC. While the individual measurements are still consistent with the NLO predictions, the central values of the various measurements are systematically higher than prediction and so large corrections beyond NLO are still allowed and even hinted at. We also note that the NLO prediction to which these measurements are compared is MCFM with the renormalization scale chosen to be $m_W$, rather than our choice of $m_{WW}$. Lowering the renormalization scale to $m_W$ serves to increase the cross section prediction and it also increases $\alpha_0'$ slightly.

We have studied a theoretical uncertainty associated with the merging of parton showers with NLO matrix elements. It arises since the merging represents a partial modeling of QCD corrections beyond NLO and thus different implementations of the merging can differ. This intrinsic uncertainty will only be reduced when true NNLO results are available. But presently it leads to a significant uncertainty in the extrapolation of a background cross section from a control region to the signal region. This could be an issue in other analyses that depend on theory for a shape of a distribution, and our approach for estimating the uncertainty could easily be carried over. It is not accounted for by varying choices of scale and PDF at the NLO level or by changing various parton shower settings. There is also the theoretical question of why the different implementations of the merging differ in the way they do, but that is beyond the scope of this work. And finally, one obvious way to reduce the uncertainty in the extrapolation is to bring the control regions and signal regions closer together, even if that means a small contamination of the control region by signal events.

\section*{Note Added}
 After the original version of this paper was posted, ATLAS released an updated $H\rightarrow WW$ analysis \cite{atlas2} based on 13 fb$^{-1}$ of $\sqrt{s}=8$ TeV data.
 
1) The event generator used for the continuum $WW$ background was changed from MC@NLO-Herwig6 to POWHEG-Pythia8. According to our Table 1 this increases the $WW$ background estimate in the 0-jet signal region by about 7\%.\footnote{This could also be seen in the original version of this paper.} We can check this by using the same procedure as described above (now using Table 4 and Fig.~(15c) of \cite{atlas2}) and thus extract $\alpha_0=0.446$ and $\alpha_0'=0.491$ from the new ATLAS analysis.\footnote{There is a new cut in the 0-jet bin but it produces a tiny effect and we ignore it.} These numbers are indeed 7\% larger than the previous numbers.

2) ATLAS notes that POWHEG-Pythia8 poorly describes the relative number of events in the 1-jet and 0-jet control regions. The 1-jet to 0-jet ratio from the data is $0.74\pm0.08$ times the POWHEG-Pythia8 prediction. We find that the Herwig++@NLO prediction is 0.71 times the POWHEG-Pythia8 prediction and so Herwig++@NLO does much better in this regard. At the same time the shapes of the $m_T$ distribution in the 0-jet control and signal regions produced by these two event generators are very similar.
 
3) We can also consider an extrapolation parameter for the 1-jet bin. According to the cuts in \cite{atlas2} we define $\alpha_1'$ in the same way as $\alpha_0'$ except that we require one $p_T>25$ GeV jet and we remove the $p_T^{\ell\ell}$ cut.\footnote{We do not implement the $b$-jet and $Z\rightarrow \tau\tau$ vetoes.} We find that Herwig++@NLO produces a value for $\alpha_1'$ which  is 12\% larger than the POWHEG-Pythia8 value.

4) The uncertainty in the extrapolation parameters that ATLAS associates with the choice of event generator (pre-shower) is 3.5\% while for parton showers and underlying event it is 4.5\%, for both the 0-jet and 1-jet bins (Table 2 of \cite{atlas2}). From our results we conclude that these numbers are underestimates of the theoretical uncertainties.

5) CMS continues \cite{cms2} to give little information about extrapolation parameters and how they are obtained. CMS uses a control region farther from the signal region ($m_{\ell\ell}>100$ GeV rather than $m_{\ell\ell}>80$ GeV) and this will increase the theoretical uncertainty of the extrapolation. The total uncertainty, which presumably includes the renormalization scale and PDF uncertainties that ATLAS estimates separately, is given as 10\%. This also appears to be an underestimate.
 
\section*{Acknowledgments}
I thank Pierre Savard for useful discussions. This work was supported in part by the Natural Science and Engineering Research Council of Canada.


\begin{thebibliography}{99}
\bibitem{diff} LHC Higgs Cross Section Working Group, S. Dittmaier, C. Mariotti, G. Passarino, and R. Tanaka (Eds.), Handbook of LHC Higgs cross sections: 2. Differential distributions, arXiv:1201.3084 [hep-ph].
\bibitem{atlas} ATLAS Collaboration, ATLAS-CONF-2012-098, 2012.
\bibitem{cms} CMS Collaboration, CMS PAS HIG-12-038, 2012.
\bibitem{diphoton} 
  S.~Catani, L.~Cieri, D.~de Florian, G.~Ferrera and M.~Grazzini,
  Phys.\ Rev.\ Lett.\  {\bf 108}, 072001 (2012)
  arXiv:1110.2375 [hep-ph].
 \bibitem{WZ} 
  F.~Campanario and S.~Sapeta,
  Phys.\ Lett.\ B {\bf 718}, 100 (2012)
  arXiv:1209.4595 [hep-ph].
\bibitem{powheg} S. Alioli, P. Nason, C. Oleari and E. Re, JHEP 1006, 043 (2010), arXiv:1002.2581 [hep-ph];
T.~Melia, P.~Nason, R.~Rontsch and G.~Zanderighi,
  JHEP {\bf 1111}, 078 (2011), arXiv:1107.5051 [hep-ph].
\bibitem{pythia} T. Sjšstrand, S. Mrenna and P. Skands, JHEP05 (2006) 026.
 \bibitem{herwig6} G. Corcella, I.G. Knowles, G. Marchesini, S. Moretti, K. Odagiri, P. Richardson, M.H. Seymour and B.R. Webber, JHEP 0101 (2001) 010, hep-ph/0011363; hep-ph/0210213. 
\bibitem{mcnlo} S. Frixione and B.R. Webber, JHEP 0206 (2002) 029, hep-ph/0204244; S. Frixione, Nucl. Phys. B410 (1993) 280Ð324; S. Frixione, F. Stoeckli P. Torrielli and B.R. Webber, JHEP 1101 (2011) 053, arXiv:1010.0568.
\bibitem{herwig} 
  M.~Bahr, S.~Gieseke, M.~A.~Gigg, D.~Grellscheid, K.~Hamilton, O.~Latunde-Dada, S.~Platzer and P.~Richardson {\it et al.},
  Eur.\ Phys.\ J.\ C {\bf 58}, 639 (2008), arXiv:0803.0883 [hep-ph];
    S.~Gieseke, D.~Grellscheid, K.~Hamilton, A.~Papaefstathiou, S.~Platzer, P.~Richardson, C.~A.~Rohr and P.~Ruzicka {\it et al.},
  arXiv:1102.1672 [hep-ph],
    K.~Arnold, L.~d'Errico, S.~Gieseke, D.~Grellscheid, K.~Hamilton, A.~Papaefstathiou, S.~Platzer and P.~Richardson {\it et al.},
  arXiv:1205.4902 [hep-ph].
\bibitem{pythia8} T. Sjšstrand, S. Mrenna and P. Skands, JHEP05 (2006) 026, Comput. Phys. Comm. 178 (2008) 852.
\bibitem{hamilton} 
  K.~Hamilton,
  JHEP {\bf 1101}, 009 (2011), arXiv:1009.5391 [hep-ph].
\bibitem{amc} \verb$http://amcatnlo.web.cern.ch/amcatnlo/$
\bibitem{mad} J.~Alwall, M.~Herquet, F.~Maltoni, O.~Mattelaer, T.~Stelzer, JHEP 1106 (2011) 128, arXiv:1106.0522. 
\bibitem{sherpa}
  T.~Gleisberg, S.~.Hoeche, F.~Krauss, M.~Schonherr, S.~Schumann, F.~Siegert and J.~Winter,
  JHEP 0902 (2009) 007, arXiv:0811.4622 [hep-ph]; T.~Gleisberg and S.~Hoeche,
  JHEP 0812 (2008) 039, arXiv:0808.3674 [hep-ph].
\bibitem{hepmc} M. Dobbs and J.B. Hansen, Comput. Phys. Commun. 134 (2001) 41.
\bibitem{mcfm} J. M. Campbell and R. K. Ellis, Phys. Rev. D 60, 113006 (1999), arXiv:hep-ph/9905386; J. M. Campbell, R. K. Ellis and C. Williams,  JHEP 1107, 018 (2011), arXiv:1105.0020 [hep-ph].
\bibitem{vbfnlo} K. Arnold, M. Bahr, G. Bozzi, F. Campanario, C. Englert, T. Figy, N. Greiner, C. Hackstein, V. Hankele, B. Jager, G. Klamke, M. Kubocz, C. Oleari, S. Platzer, S. Prestel, M. Worek, D. Zeppenfeld,
Comput. Phys. Commun. 180, 1661 (2009), arXiv:0811.4559  [hep-ph]; K. Arnold, J. Bellm, G. Bozzi, M. Brieg, F. Campanario, C. Englert, B. Feigl, J. Frank, T. Figy, F. Geyer, C. Hackstein, V. Hankele, B. Jager, M. Kerner, M. Kubocz, C. Oleari, S. Palmer, S. Platzer, M. Rauch, H. Rzehak, F. Schissler, O. Schlimpert, M. Spannowsky, M. Worek, D. Zeppenfeld, arXiv:1107.4038 [hep-ph]; K. Arnold, J. Bellm, G. Bozzi, F. Campanario, C. Englert, B. Feigl, J. Frank, T. Figy, B. Jager, M. Kerner, M. Kubocz, C. Oleari, S. Palmer, M. Rauch, H. Rzehak, F. Schissler, O. Schlimpert, M. Spannowsky, D. Zeppenfeld, arXiv:1207.4975 [hep-ph].
\bibitem{delphes} S. Ovyn, X. Rouby, V. Lemaitre, ``Delphes, a framework for fast simulation of a generic collider experiment'', arXiv:0903.2225 [hep-ph].
\bibitem{fastjet}
  M.~Cacciari, G.~P.~Salam and G.~Soyez,
  Eur.\ Phys.\ J.\ C {\bf 72} (2012) 1896, arXiv:1111.6097 [hep-ph].
 \bibitem{mada} F. Maltoni and R. Frederix.
\bibitem{atlas2}ATLAS Collaboration, ATLAS-CONF-2012-158, 2012.
\bibitem{cms2}CMS Collaboration, CMS PAS HIG-12-042, 2012.


\end{thebibliography}
\end{document}